**Direct counterfactual quantum communication protocol beyond single photon source**

Zheng-Hong Li,[1,2,3,*] Shang-Yue Feng,[1] M. Al-Amri,[4,5] and M. Suhail Zubairy[4]

1 Department of Physics, Shanghai University, Shanghai 200444, China

2 Zhejiang Province Key Laboratory of Quantum Technology and Device, Zhejiang University, Hangzhou 310027, China

3 Shanghai Key Laboratory of High Temperature Superconductors, Shanghai University, Shanghai 200444, China

4 Institute for Quantum Science and Engineering (IQSE) and Department of Physics and Astronomy, Texas A&M University, College Station, Texas 77843-4242, USA

5 NCQOQI, KACST, P.O.Box 6086, Riyadh 11442, Saudi Arabia

The direct counterfactual quantum communication protocol involving double chained Mach-Zehnder interferometers requires a single photon input. Here, we show that even with multiphoton light sources, including a strong coherent light source as inputs, the counterfactual communication can be achieved with success probability approaching unity in the ideal asymptotic limit. The path evolution of multiple photons is non-locally controlled. Thus, information is transmitted without any photons, or any other auxiliary information carriers, appearing in the public transmission channel. The effect is quantum since quantum measurements are essential requirement for this protocol. Furthermore, a modified scheme is proposed in which number of interferometers is reduced.

## I. Introduction

Measurement plays a unique role in quantum mechanics and leads to many counterintuitive quantum phenomena. One of them is the direct counterfactual quantum communication SLAZ protocol originally proposed by Salih, Li, Al-Amri, and Zubairy [1]. This protocol is inspired by interaction free measurement [2,3] along with the use of quantum Zeno effect [4,5]. It can transmit one bit of information remotely without any physical particles traveling between two communicating parties Alice and Bob. The probability of obtaining the information correctly in the ideal asymptotic limit approaches 100% [1]. Although the SLAZ protocol requires time to transmit information, the non-local quantum effect that it presents in it has triggered intense research interest in fundamental research [6-17]. As a non-local quantum control method for a single photon, the SLAZ protocol has led to new ideas about entanglement preparation [18-21], and quantum teleportation [22-25]. Furthermore, further research has been done on topics such as quantum gates [26-29], ghost imaging [30], and so on [31,32]. In addition, there have been proof-of-principle experiments that were carried out. In 2017, the research groups of Pan and Zhu independently published verification experiments [33,34].

In the SLAZ protocol, a single photon forms a three-path superposition state, one of which is the public transmission channel connecting Alice and Bob. The other two paths are confined to Alice's end. The basic idea is to force the photon state to collapse to only these two paths through multiple measurements of the photon in the transmission channel, so that information can be transmitted without the photon appearing in the transmission channel. When multiple photons are considered, however, the situation is no longer simple. Under ordinary conditions, multiple photons cannot act collectively (just like a single photon), resulting in forming a photon number distribution in the three paths. In other words, the light field is conventionally described by light intensity distribution rather than the probabilistic description. Under the description of light intensity, there is always non-zero photon state components in the public transmission channel. This leads to the current mainstream view that multiphoton light sources, especially strong coherent light sources, cannot achieve direct counterfactual quantum communication [33,34].

In this work, we show that the above scenario is not always the case. The aforementioned non-local quantum effect can still be true for multiphoton light sources. It is possible to non-locally control the path evolution of multiple photons, including strong coherent light field, so that information can still be transmitted counterfactually. It should be emphasized that, in this article, we focus on whether the non-local quantum effects can theoretically occur, and hence unlock the physics behind it. Therefore, our proof only considers ideal conditions, i.e., equipments and environments. The reason why we write article from an information transmission perspective is to give the reader a simple and intuitive example, and to illustrate the relevance to ref. [1]. Hence, we simply consider the model in which Bob transmits 1-bit information to Alice, and our discussion does not involve those recently developed protocols with modified definitions and criteria [13-17]. On this basis, we first study the SLAZ protocol with multi-photon source instead of a single photon source, and hence analyze their similarities and differences. Our study shows the conditions for counterfactual communication are more stringent, requiring more resources. Nonetheless, we then propose a modified scheme, which can greatly reduce the resources that are required compared with the SLAZ protocol despite using the same multi-photon source.

## II. Basic scheme and calculation method

In Fig. 1(a), $S_L$ stands for light source, $MR$ stands for mirror, $BS$ stands for beam-splitter, and $D$ stands for detector. The scheme has the structure of double chained Mach-Zehnder interferometers. There are $M$ $BS_M$ constituting the outer chain, and the lower arm of each interferometer in the outer chain is embedded with an inner chain composed of $N$ $BS_N$. According to Ref. [1], we call $M(N)$ the outer (inner) cycle number, and $T = MN$ the total cycle number. For the description of the function of the $BS_{M(N)}$, we divide Fig. 1(a) into three zones. We assume that the photon state $|v_0, v_1, v_2\rangle = \prod_{j=0,1,2}[(a_j^\dagger)^{v_j}/\sqrt{v_j!}]|0,0,0\rangle$ represents $v_j$ photons in Zone $j$, and $a_j^\dagger$ is the corresponding creation operator. Then, the function of $BS_{M(N)}$ can be described as

$$a_{0(1)}^\dagger \to a_{0(1)}^\dagger \cos\theta_{M(N)} + a_{1(2)}^\dagger \sin\theta_{M(N)},$$
$$a_{1(2)}^\dagger \to a_{1(2)}^\dagger \cos\theta_{M(N)} - a_{0(1)}^\dagger \sin\theta_{M(N)}, \quad (1)$$

where $\theta_{M(N)} = \pi/2M(N)$ and $\cos^2\theta_{M(N)}$ represents the reflectivity of $BS_{M(N)}$. In the lower arms of the inner chains, the white belt stands for the public transmission channel connecting Alice and Bob. At Bob's end, the detector $D_{2B}$ is activated for his signal $s = 1$, and deactivated (becomes transparent) for signal $s = 0$. At Alice's end, $D_0$ and $D_1$ are used to receive Bob's signal.

Before we consider the complete scheme, we first focus on the inner chain. The evolution of photons in the outer chain obeys the same laws. Let us assume that the input of the inner chain is a Fock state $|0, v, 0\rangle$. When $s = 0$, it is easy to see that, after $n$ $BS_N$, the photon state is $(a_1^\dagger \cos n\theta_N + a_2^\dagger \sin n\theta_N)^v/\sqrt{v!}|0,0,0\rangle$. When $n = N$, all photons entering the inner chain are routed to $D_{2A}$ side. However, when $s = 1$, interference in the chain is interrupted by $D_{2B}$'s continuous measurements. Unless otherwise specified, in this work we only require the detector to be able to distinguish between the vacuum state ($|0\rangle$) and the non-zero photon number state. If $D_{2B}$ cannot find any photons in the transmission channel after $N$ measurements, the photon state becomes $(a_1^\dagger \cos^N \theta_N)^v / \sqrt{v!} |0,0,0\rangle$, where those terms containing $a_2^\dagger$ are eliminated since the photon state in Zone 2 must collapse to $|0\rangle$. In addition, it is worth pointing out that when $N \gg 1$, $\cos^N \theta_N \approx 1 - \pi^2/8N$, and this approximation also works for $M$.

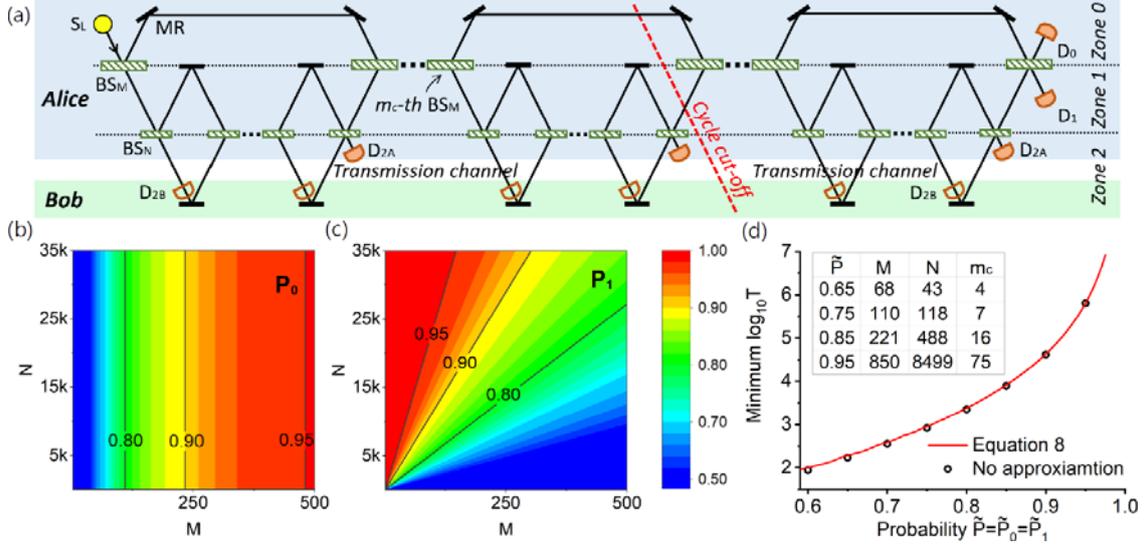

FIG. 1 (a) Basic scheme based on double chained Mach-Zehnder interferometers. When the input state is a coherent state with average photon number $|\alpha|^2 = 10$, we plot (b) the probability ($P_0$) that only $D_0$ clicks when $s = 0$ and (c) the probability ($P_1$) that only $D_1$ clicks when $s = 1$ for different $M$ and $N$. In the modified scheme, the communication is cut off at the red dashed line where $D_0$ and $D_1$ are repositioned. In order to analyze the needed resources described by the total cycle number $T$ in the modified scheme, we plot (d) the minimum $\log_{10} T$ for different $\tilde{P} = \tilde{P}_0 = \tilde{P}_1$, when $|\alpha|^2 = 200$, where $\tilde{P}_s$ represents the probability that no photon enters the transmission channel and $D_s$ detects at least one photon for Bob's signal $s$. The red solid curve is based on Eq. (8), and circle points come from the complete simulation without any approximation.

### III. Theory and demonstration

In this section, we prove that even if Alice uses a multiphoton light source, with the help of multiple measurements, it is possible to transmit one bit of information directly from one communicator (Bob) to the other (Alice) such that, during the entire information transmission process, the photon state at Bob's end is a vacuum state. The probability of this happening tends towards 100% as system resource (number of interferometers) increases.

Our calculations are based on the dynamical evolution of Alice's photons. The basic idea is as follows. Taking the case that Bob's signal $s = 1$ as an example, we notice that photon exchange between Zone 1 and the public transmission channel occurs only when Alice's photons pass through $BS_N$. Therefore, after each pass, $D_{2B}$ is used to ensure that no photons appear in the transmission channel due to measurement, which causes the quantum state to collapse. After the measurement, if $D_{2B}$ does not click, the collapsed photon state becomes the initial state for the next dynamical evolution. Eventually, we can obtain $P_1$, which is the probability of no detectors being triggered except for $D_1$ during the entire dynamical evolution process. Since $D_{2B}$ is always silent, this ensures that Bob's state is always in a vacuum. As for the case where Bob's signal is 0, the situation is similar, except that we use $D_{2A}$ to ensure that there are no photons in the entire Zone 1 and 2 (including the transmission channel). In this case, the corresponding probability is $P_0$, i.e., only $D_0$ clicks. We show that the two final states of Alice's photons according to Bob's different signals are orthogonal. Therefore, Alice can receive Bob's information correctly. More importantly, we show that the probabilities $P_1$ and $P_0$ can both tend towards 1. In such extreme case, Alice's photons are manipulated non-locally and receive Bob's information correctly. We consider this as the realization of direct counterfactual quantum communication.

### A. Fock state input case

Consider the situation where the input photon state of Fig. 1(a) is a Fock state $|v,0,0\rangle$. According to the discussion in Section II, when $s = 0$, any photons entering the inner chain must be routed to $D_{2A}$, so the inner chain and $D_{2A}$ can be treated as a combined detector. Consequently, after $m$ $BS_M$, if $D_{2A}$ does not click, the photon state is $|\phi_0^m(v)\rangle = \cos^{(m-1)v}\theta_M \left(a_0^\dagger \cos\theta_M + a_1^\dagger \sin\theta_M\right)^v / \sqrt{v!} \, |0,0,0\rangle$. When $M \gg 1$, it is approximately given by

$$|\phi_0^M(v)\rangle = \frac{1}{\sqrt{v!}}\left[\left(1 - \frac{\pi^2}{8M}\right)a_0^\dagger + \frac{\pi}{2M}a_1^\dagger\right]^v |0,0,0\rangle. \qquad (2)$$

Regarding the case $s = 1$, the photons entering the transmission channel are measured by $D_{2B}$. When $N$ is large, the photons have a large probability of being retained on the upper side of the inner chain. However, as long as $N$ is finite, the multiple measurements must cause photon loss and lead to the imbalance of the two arms of the interferometer in the outer chain. To account for this, we do power series expansion and discard all the second order and higher order terms of $1/N$. Consequently, after $m$ $BS_M$, if $D_{2B}$ does not click, the photon state is $|\phi_1^m(v)\rangle = \left[a_0^\dagger(\cos m\theta_M + A_m) + a_1^\dagger(\sin m\theta_M - B_m)\right]^v / \sqrt{v!} \, |0,0,0\rangle$ where $A_m = (\pi^2/8N)\sum_{m'=1}^{m-1} \sin m'\theta_M \sin(m - m')\theta_M$ and $B_m = (\pi^2/8N)\sum_{m'=1}^{m-1} \sin m'\theta_M \cos(m - m')\theta_M$. When $m = M$, the summations can be replaced by integrations [35], which leads to $A_M \approx \pi M/8N$ and $B_M \approx M\pi^2/16N$. Therefore, for $M \gg 1$, we have,

$$|\phi_1^M(v)\rangle = \frac{1}{\sqrt{v!}}\left[\frac{\pi M}{8N}a_0^\dagger + \left(1 - \frac{\pi^2 M}{16N}\right)a_1^\dagger\right]^v |0,0,0\rangle. \qquad (3)$$

More details on the calculations of Eqs. (2) and (3) can be found in Appendix A.

Based on Eqs. (2) and (3), we calculate the probability $P_s$ that $D_s$ detects at least one photon and no other detector clicks when Bob's signal is $s$. We emphasize that any photons entering the transmission channel (Zone 2) must cause $D_{2A}$ or $D_{2B}$ to be triggered. Therefore, if both $P_0$ and $P_1$ approach 1, it means that one bit of information can be transmitted directly without any photon entering the transmission channel, which leads to the direct counterfactual quantum communication. In the Fock state input case, it is easy to see that $P_{0(1)}$ is equal to the probability of Alice detecting the final state $|v,0,0\rangle$ ($|0,v,0\rangle$). Then, under the condition that $N \gg vM \gg v^2$, we have $P_0 \approx 1 - \pi^2 v/4M$ and $P_1 \approx 1 - \pi^2 Mv/8N$, which approach unity. Consequently, the counterfactual communication can be realized with a Fock state input. Obviously, the single photon case proposed in Ref. [1] is a special case.

### B. Arbitrary photon statistics input case

Consider the initial photon state $\sum_{v=0}^{\infty} c_v|v,0,0\rangle$, where $\sum_{v=0}^{\infty}|c_v|^2 = 1$. Since for a specific Fock state $|v,0,0\rangle$, the total photon number $v$ is conserved throughout the communication process if both $D_{2A}$ and $D_{2B}$ do not click, the dynamic evolution of Fock states with different total photon numbers can not interfere with each other. Consequently, we obtain

$$P_0 = \sum_{v=1}^{\infty}|c_v|^2 |\langle v,0,0|\phi_0^M(v)\rangle|^2 = \sum_{v=1}^{\infty}|c_v|^2 \left(1 - \frac{\pi^2}{8M}\right)^{2v},$$

$$P_1 = \sum_{v=1}^{\infty}|c_v|^2 |\langle 0,v,0|\phi_1^M(v)\rangle|^2 = \sum_{v=1}^{\infty}|c_v|^2 \left(1 - \frac{\pi^2 M}{16N}\right)^{2v}. \quad (4)$$

It is worth mentioning that the summation is from $v=1$, since initial state $|0,0,0\rangle$ does not contribute to transmission of information. Accordingly, if we want $P_s \to 1$, the first requirement is that $|c_0|^2 \to 0$ (This ensures that Alice's two final states are orthogonal). In addition to the above requirement, a general condition for $P_s \to 1$ is given as follows. Consider the case where the light field has a cutoff $v_c$ satisfying $\sum_{v_c}^{\infty}|c_v|^2 v \to 0$, which means that the photon number states with $v > v_c$ have no contribution to the average photon number $\bar{v} = \sum_{v=0}^{\infty}|c_v|^2 v$. Then, when $N \gg v_c M \gg v_c^2$ (when considering a specific photon number distribution, this condition can be optimized, as shown in Appendix B), we can obtain

$$P_0 \approx 1 - \pi^2 \bar{v}/4M, \qquad P_1 \approx 1 - \pi^2 \bar{v} M/8N. \quad (5)$$

More details are in Appendix C. Consequently, both $P_0$ and $P_1$ can approach unity so that the direct counterfactual quantum communication can be achieved.

### IV. Discussions
#### A. Finite $M$ and $N$ scenario, and the role of quantum measurement

We have shown that using the optical scheme proposed in the SLAZ protocol, the counterfactual communication can be achieved with multiphoton sources, and its success probability approaches unity in the ideal asymptotic limit. However, in practice, $M$ and $N$ are finite. Under this condition, there is a difference between a Fock state input and an input with arbitrary photon statistics. For the Fock state case, when Alice collects all $v$ photons, she knows that no photons entered the transmission channel. In other words, the measurement results of $D_{2A}$ and $D_{2B}$ are not necessary. However, for a more general input, only the result of local measurements at Alice's end cannot help her reach the same conclusion. For example, if the input state is $c_1|1,0,0\rangle + c_2|2,0,0\rangle$ and Bob's $D_{2B}$ detects one photon, this may result in the final state $|1,0,1\rangle$. Unless Alice has Bob's measurement results, she cannot distinguish between $|1,0,0\rangle$ and $|1,0,1\rangle$. To ensure that Bob delivers results to Alice without transferring any physical particles, we require that Bob makes announcements only when a photon is detected (It is worth emphasizing that in this case, the counterfactual communication has already failed), and to remain silent at other times. *The key here is that Bob has the ability to deliver both signals 0 and 1 to Alice* [7]*, while remaining silent, and the probability of Bob being silent approaches unity as M and N increase.* Evidently, since Bob's quantum measurement results are an essential part, the effect is manifestly quantum.

As a result, having the Fock state as an input including the special case when there is only a single photon (Ref.[1]) is an extreme situation of the direct counterfactual quantum communication. It is worth mentioning that such a light source has quantum characteristics. On the contrary, the other extreme case is the coherent light source case, where the counterfactual effect is only determined by multiple quantum measurements. Consider a coherent state input $|\alpha, 0\rangle$ passing through a $BS$. According to Eq. (1), the outputs $|\alpha \cos \theta, \alpha \sin \theta\rangle$ are still coherent states. However, once we perform a measurement on one side of the $BS$, the photon state has a probability of $\exp(-|\alpha|^2 \sin^2 \theta)$ to collapse to $|\alpha \cos \theta, 0\rangle$. Regardless of the average number of photons, $|\alpha|^2$, of the input, this probability can approach unity by reducing the transmittance ($\sin^2 \theta$) of the $BS$. As a result, in the coherent state input case, clicks at $D_0$ and $D_1$ as well as the measurement results of $D_{2A}$ and $D_{2B}$, are essential to ensure counterfactuality.

### B. Resources needed for the single-photon case and the coherent light case

To fully demonstrate the non-local effect due to the critical role of multiple measurements, we now focus on the strong coherent input state $|\alpha, 0, 0\rangle = \exp(-|\alpha|^2/2) \sum_{v=0}^{\infty} \alpha^v / \sqrt{v!} |v, 0, 0\rangle$, which satisfies $|c_0|^2 = \exp(-|\alpha|^2) \approx 0$. In Appendix B, we obtain the same result as Eq. (5). However, the condition for counterfactual communication changes to $N \gg \bar{v} M \gg \bar{v}^2$. To verify our results, in Fig. 1(b, c) we show the numerical simulation results of $P_0$ and $P_1$ for $|\alpha|^2 = 10$ for different $M$ and $N$. For example, when $M = 250$ and $N = 35000$, the numerical results are $P_0 = 0.906$ and $P_1 = 0.916$, while the approximate results given by Eq. (5) are $P_0 = 0.901$ and $P_1 = 0.912$. All these results are in good agreement with each other.

Based on Eq. (5), we can analyze the resources that are required and compare with the single photon case. We note that there are $M - 1$ inner chains in the entire optical structure, and $N - 1$ interferometers in each inner chain as shown in Fig.1(a). Alice's photons pass through these interferometers one by one, thus we can use the total number of cycle $T = MN$ to describe the complexity of the optical system and the required resources. Furthermore, $T$ can also describe the time taken for information extraction if we assume that the optical distance of each interferometer is consistent.

Based on the above definition, we compare the coherent light case and the single photon case of the SLAZ protocol. Regarding the single photon case, we assume that its outer (inner) cycle number is $M'$ ($N'$). Accordingly, the probabilities of $D_0$ and $D_1$ detecting the single photon are $P'_0 = 1 - \pi^2/4M'$ and $P'_1 = 1 - \pi^2 M'/8N'$, respectively. Regarding the case of coherent input state $|\alpha\rangle$, if we want $P_s = P'_s$, the requirement for the total cycle number is $MN = |\alpha|^6 M'N'$. Obviously, if the effect of the average photon number $|\alpha|^2$ is ignored and the experiment is performed with only the parameters in the single-photon case, no counterfactual effect can be observed. This is exactly what happened in Ref. [34]. In that experiment, a counterfactual-like effect is reported, in which only the intensity distribution result is observed.

Going back to the current work, unfortunately, there is a pitfall with the increase of $|\alpha|^2$, which leads the needed resources to be greatly increased. To overcome this, we next present a modified scheme, which can significantly reduce resources comparing with SLAZ protocol using the same light source.

### V. Modified scheme

Here we propose a modified scheme for strong coherent input state $|\alpha\rangle$. The corresponding calculation details can be found in Appendix D. In the original scheme, when $s = 1$, photons are routed from Zone 0 to Zone 1, but when $s = 0$, at the end of each inner chain, the photon number in Zone 1 is exactly zero. From the perspective of information transmission, there is no need to wait for all the photons to move to Zone 1, and then confirm that Bob's signal is $s = 1$. Based on this idea, we reposition $D_0$ and $D_1$ at the red dashed line, after the $m_c$-th inner chain in Fig. 1(a), and end the communication process there. Here, we emphasize that, in this set-up, the number of $BS_M$ is determined by $m_c$, but the reflectivity of $BS_M$ is still determined by $M$. As for $BS_N$, both the reflectivity and the total number of $BS_N$ in each inner chain are determined by $N$. Therefore, the total cycle number required in the modified scheme is $T = m_c N$. Apparently, the determination of an appropriate value of $m_c$ is critical to minimize $T$, which will be discussed below.

Based on previous discussions, the photon states on the red dashed line are $|\widetilde{\Phi}_0^{m_c}\rangle = \exp\left[-\frac{1}{2}|\alpha|^2(1 - \cos^{2m_c}\theta_M)\right]|\alpha\cos^{m_c}\theta_M, 0, 0\rangle$ for $s = 0$, and $|\widetilde{\Phi}_1^{m_c}\rangle = \exp\left[-|\alpha|^2\frac{\pi^2}{8N}\sum_{m'=1}^{m_c}\sin^2 m'\theta_M\right]\left|\alpha(\cos m_c\theta_M + A_{m_c}), \alpha\left[\left(1 - \frac{\pi^2}{8N}\right)\sin m_c\theta_M - B_{m_c}\right], 0\right\rangle$ for $s = 1$. In the following, we calculate the probability $\widetilde{P}_s$ that $D_{2A}$ and $D_{2B}$ do not click, and $D_s$ detects at least one photon for Bob's signal $s$. Obviously, once $D_1$ is triggered, Bob's signal must be $s = 1$. If the communication system can guarantee that $\widetilde{P}_1 \to 1$, then the silence of $D_1$ indicates that Bob's signal is $s = 0$ according to $|\widetilde{\Phi}_0^{m_c}\rangle$. As a result, the direct counterfactual quantum communication can be achieved when both $\widetilde{P}_0$ and $\widetilde{P}_1$ approach unity.

To get the expression of $\widetilde{P}_s$, we define the average photon number expected in Zone 1 at the end of communication process when s = 1 as $\bar{k}$. If the probability of any photon leaking out of Alice's device is 0, we have $\bar{k} \approx (|\alpha|m_c\theta_M)^2$ when $\bar{k} \ll |\alpha|^2$ (See Appendix D). Based on this approximation, we have

$$\widetilde{P}_0 = \exp\left(-|\alpha|^2\frac{m_c\pi^2}{4M^2}\right), \tag{6}$$

$$\widetilde{P}_1 = \exp\left[\frac{\pi^2(m_c + 1)(2m_c + 1)\ln\widetilde{P}_0}{24N}\right]\left\{1 - \exp\left[m_c\ln\widetilde{P}_0\left(1 - \pi^2\frac{1 + m_c}{16N}\right)^2\right]\right\}. \tag{7}$$

In Eq. (7), we have used $M = \sqrt{-|\alpha|^2 m_c \pi^2 / 4\ln\widetilde{P}_0}$ which follows from Eq. (6). In addition, Eq. (6) also indicates that $m_c = -\bar{k}/\ln\widetilde{P}_0$. Since the probability of not triggering $D_1$ tends towards zero when $\bar{k}$ is large, i.e., $\exp(-\bar{k}) = \exp(m_c\ln\widetilde{P}_0) \approx 0$, it is not difficult to obtain $N = \pi^2\bar{k}^2/12\ln\widetilde{P}_0\ln\widetilde{P}_1$ from Eq. (7). This means that for given values of $|\alpha|^2$ and $\bar{k}$, even if $\widetilde{P}_0$ and $\widetilde{P}_1$ are very close to 1, we can still have the corresponding parameters $M$, $N$ and $m_c$. Moreover, if we compare the single photon case by assuming that $\widetilde{P}_s = P'_s$, we can get that $T = 32\bar{k}^3 N'M'/3\pi^4 \approx \bar{k}^3 N'M'/10$. Note that for the SLAZ protocol using the same light source, the total cycle number is $|\alpha|^6 M'N'$. Since $\bar{k} \ll |\alpha|^2$, the modified scheme is more resource-efficient than the original SLAZ protocol with coherent input state $|\alpha\rangle$ (See Appendix D).

In the above discussion, the value of $\bar{k}$ is given in advance. However, Eq. (7) implies that there is an optimal $\bar{k}$ (or $m_c$) for the minimum $T$ when $\widetilde{P}_0$, $\widetilde{P}_1$ and $|\alpha|^2$ are given. To get the minimum $T$, we derive an expression for $N$ in terms of $m_c$ from Eq. (7). We note that the first exponential term in Eq. (7) comes from the probability that no photon enters the transmission channel, which approaches unity when $N \gg m_c^2$. Under this condition, except the term $\exp(m_c\ln\widetilde{P}_0)$, other exponential terms in Eq. (7) can be replaced by their first-order power series expansion of $1/N$, leading to

$$N = \frac{\pi^2(m_c + 1)\big[(2m_c + 1) + (m_c - 1)\exp(m_c \ln \tilde{P}_0)\big] \ln \tilde{P}_0}{24\big[\tilde{P}_1 + \exp(m_c \ln \tilde{P}_0) - 1\big]}. \tag{8}$$

It follows from Eq. (8) that, for given values of $\tilde{P}_0$, $\tilde{P}_1$ and $|\alpha|^2$, we only need to scan $m_c$ to get the minimum $T$. Based on this idea, in Fig. 1(d), we plot the red solid curve of minimum $\log_{10} T$ versus $\tilde{P} = \tilde{P}_0 = \tilde{P}_1$ with $|\alpha|^2 = 200$. As a comparison, the circle points are derived from the numerical simulation without any approximation. As shown in the figure, the two curves almost overlap.

In the table of Fig. 1(d), we give the values of $M$, $N$ and $m_c$ when $T$ is the minimum for some values of $\tilde{P}$. Our numerical simulation also shows that for $\tilde{P} = 0.5$ and $|\alpha|^2 = 200$, $T$ is the minimum when $m_c = 2$, $N = 14$, $M = 38$. We mention that this set of parameters is close to the current experimental conditions in Ref.[34], except that the reflectivity of $BS_M$ has changed drastically. Without this change, the probability of detecting photon in the transmission channel is almost unity. Due to above comparison with Ref [34], we expect that the modified scheme may be beneficial for future proof of principle experiments. Another change worth mentioning compared to Ref.[34] is the type of the detector. It is not difficult to find that when $s = 0$, in the last inner interferometer of the first outer interferometer, the average number of photons appearing in the transmission channel reaches its maximum value in our modified scheme. It is $|\alpha|^2 \sin^2 \theta_M \approx 0.34$ when using the parameters mentioned above. As for $s = 1$, the maximum value appears in the first inner interferometer of the last outer interferometer, which is $|\alpha|^2 \sin^2 m_c\theta_M \sin^2 \theta_N \approx 0.017$. Since our protocol only requires that the detector to distinguish between vacuum state and non-zero photon state, a single-photon detector is sufficient for this task. Consequently, compared with the experiment in Ref.[34], the main changes are only the adjustment of the reflectivity of $BS_M$ and the use of single-photon detectors.

### VI. Conclusion

We have demonstrated that single photon source is not a necessary condition for direct counterfactual quantum communication, that is, when Bob's photon state continues to be in a vacuum state, even with a multiphoton light source, he can transmit one bit of information to Alice without the help of any physical carriers, and the success probability is close to 100% in the ideal asymptotic limit. We show that multiple quantum measurements play a critical role in the above process. Moreover, we propose a modified scheme that can reduce the needed resources compared with the SLAZ protocol albeit using the same multi-photon source.

### Acknowledgements:
This work is supported by National Natural Science Foundation of China (NSFC) (No. 11704241), a grant from the King Abdulaziz City for Science and Technology (KACST), and an NPRP grant (NPRP135-0205-200258) from the Qatar National Research Foundation (QNRF).

### APPENDIX A: Calculations for Fock state as an input state

According to Fig.1(a) in the main text, $a_0$, $a_1$ and $a_2$ represent the annihilation operators of the light field in Zones 0, 1 and 2, respectively. Based on this, the function of the $BS_{M(N)}$ can be described as $a^\dagger_{0(1)} \to a^\dagger_{0(1)} \cos\theta_{M(N)} + a^\dagger_{1(2)} \sin\theta_{M(N)}$ and $a^\dagger_{1(2)} \to a^\dagger_{1(2)} \cos\theta_{M(N)} - a^\dagger_{0(1)} \sin\theta_{M(N)}$, where $\theta_M = \pi/2M$ and $\theta_N = \pi/2N$. Now, we consider the initial photon state

$$|v,0,0\rangle = \frac{1}{\sqrt{v!}}(a_0^\dagger)^v|0,0,0\rangle, \tag{A1}$$

which represents that $v$ photons appear in Zone 0, while no photons appear in Zones 1 and 2. After the first $BS_M$,

$$|\phi_0^1\rangle = |\phi_1^1\rangle = \frac{1}{\sqrt{v!}}\left(a_0^\dagger \cos\theta_M + a_1^\dagger \sin\theta_M\right)^v|0,0,0\rangle, \tag{A2}$$

where the superscript of $|\phi_s^m\rangle$ represents the $m$-th outer interferometer, while the subscript represents Bob's signal $s$.

In the first inner chain, after the first $BS_N$, the photon state becomes

$$\frac{1}{\sqrt{v!}}\left[a_0^\dagger \cos\theta_M + \sin\theta_M\left(a_1^\dagger \cos\theta_N + a_2^\dagger \sin\theta_N\right)\right]^v|0,0,0\rangle. \tag{A3}$$

**A.1 Bob deactivates his detector $D_{2B}$, i.e., his signal is $s = 0$.**

In the first inner chain, after $n$-th $BS_N$, the photon state becomes

$$\frac{1}{\sqrt{v!}}\left[a_0^\dagger \cos\theta_M + \sin\theta_M\left(a_1^\dagger \cos n\theta_N + a_2^\dagger \sin n\theta_N\right)\right]^v|0,0,0\rangle. \tag{A4}$$

When $n = N$, we have

$$\frac{1}{\sqrt{v!}}\left(a_0^\dagger \cos\theta_M + a_2^\dagger \sin\theta_M\right)^v|0,0,0\rangle$$

$$= \frac{1}{\sqrt{v!}}\sum_{j=0}^{v} C_v^j \left(a_0^\dagger \cos\theta_M\right)^{v-j}\left(a_2^\dagger \sin\theta_M\right)^j |0,0,0\rangle$$

$$= \frac{1}{\sqrt{v!}}\sum_{j=0}^{v} C_v^j \cos^{v-j}\theta_M \sin^j\theta_M \sqrt{(v-j)!\,j!}\,|v-j,0,j\rangle. \tag{A5}$$

Right now, $D_{2A}$ measures the photons appearing in Zone 2. If it does not click, the photon state collapses to

$$\frac{1}{\sqrt{v!}}\cos^v\theta_M\left(a_0^\dagger\right)^v|0,0,0\rangle, \tag{A6}$$

i.e., all terms with $a_2^\dagger$ are eliminated. Here, we emphasize that we do not perform normalization to facilitate probability calculations. After the second $BS_M$, the photon state becomes

$$|\phi_0^2\rangle = \frac{1}{\sqrt{v!}}\cos^v\theta_M\left(a_0^\dagger \cos\theta_M + a_1^\dagger \sin\theta_M\right)^v|0,0,0\rangle. \tag{A7}$$

The above process is repeated in each outer interferometer. After $m$-th $BS_M$, we have

$$|\phi_0^m\rangle = \frac{1}{\sqrt{v!}}\cos^{(m-1)v}\theta_M\left(a_0^\dagger \cos\theta_M + a_1^\dagger \sin\theta_M\right)^v|0,0,0\rangle. \tag{A8}$$

When $m = M$, we have

$$|\phi_0^M\rangle = \frac{1}{\sqrt{v!}} \cos^{(M-1)v} \theta_M \left(a_0^\dagger \cos\theta_M + a_1^\dagger \sin\theta_M\right)^v |0,0,0\rangle$$
$$\approx \frac{1}{\sqrt{v!}} \left[a_0^\dagger \left(1 - \frac{\pi^2}{8M}\right) + a_1^\dagger \frac{\pi}{2M}\right]^v |0,0,0\rangle. \tag{A9}$$

*This is Eq. (2) in the main text.* In the second line, we have used the approximation that

$$\cos^M \theta_M \approx 1 - \frac{\pi^2}{8M} \tag{A10}$$

and

$$\sin\theta_M \approx \frac{\pi}{2M}. \tag{A11}$$

We neglect all second or higher order terms of $1/M$.

Now we can calculate the probability that $D_0$ and only $D_0$ clicks. Under such condition, it is easy to see that the photon state has to be $|v,0,0\rangle$. Then, we have

$$P_0 = \cos^{2Mv} \theta_M \approx \left(1 - \frac{\pi^2}{8M}\right)^{2v} \approx 1 - \frac{\pi^2 v}{4M} \tag{A12}$$

To get the third equation, we require that $M \gg v$.

### A.2 Bob activates his detector $D_{2B}$, i.e., his signal is $s = 1$.

Based on Eq.(A3), we consider the influence of Bob's measurement. If $D_{2B}$ does not click, we have

$$\frac{1}{\sqrt{v!}} \left(a_0^\dagger \cos\theta_M + a_1^\dagger \sin\theta_M \cos\theta_N\right)^v |0,0,0\rangle. \tag{A13}$$

After $N$ $BS_N$, we have

$$\frac{1}{\sqrt{v!}} \left(a_0^\dagger \cos\theta_M + a_1^\dagger \sin\theta_M \cos^N \theta_N\right)^v |0,0,0\rangle$$
$$\approx \frac{1}{\sqrt{v!}} \left[\left(a_0^\dagger \cos\theta_M + a_1^\dagger \sin\theta_M\right) + a_1^\dagger \left(-\frac{\pi^2}{8N}\right) \sin\theta_M\right]^v |0,0,0\rangle. \tag{A14}$$

Here, in the second line, we have used the approximation

$$\cos^N \theta_N \approx 1 - \frac{\pi^2}{8N}, \tag{A15}$$

which requires $N \gg 1$.

After the second $BS_M$, we have

$$|\phi_1^2\rangle = \frac{1}{\sqrt{v!}} \left[a_0^\dagger \cos 2\theta_M + a_1^\dagger \sin 2\theta_M\right.$$
$$\left. + \left(-\frac{\pi^2}{8N}\right) \sin\theta_M \left(a_1^\dagger \cos\theta_M - a_0^\dagger \sin\theta_M\right)\right]^v |0,0,0\rangle. \tag{A16}$$

The above process is repeated in the next outer interferometers. After the third $BS_M$, we have

$$|\phi_1^3\rangle = \frac{1}{\sqrt{v!}} \left[\left(a_0^\dagger \cos 3\theta_M + a_1^\dagger \sin 3\theta_M\right)\right.$$

$$+ \left(-\frac{\pi^2}{8N}\right) \sin\theta_M \left(a_1^\dagger \cos 2\theta_M - a_0^\dagger \sin 2\theta_M\right)$$

$$+ \left(-\frac{\pi^2}{8N}\right) \sin 2\theta_M \left(a_1^\dagger \cos\theta_M - a_0^\dagger \sin\theta_M\right)\Big]^v |0,0,0\rangle. \tag{A17}$$

Here we continue to use the approximation shown in Eq. (A15) and we have also neglected all second or higher order terms of $1/N$.

Similarly, after $m$-th $BS_M$, we have

$$|\phi_1^m\rangle = \frac{1}{\sqrt{v!}}\{(a_0^\dagger \cos m\theta_M + a_1^\dagger \sin m\theta_M)$$

$$+ \left(-\frac{\pi^2}{8N}\right) \sum_{m'=1}^{m-1} \sin m'\theta_M \left[a_1^\dagger \cos(m-m')\theta_M - a_0^\dagger \sin(m-m')\theta_M\right]\}^v |0,0,0\rangle. \tag{A18}$$

Since $\theta_M = \pi/2M$, when $m = M$, we have

$$|\phi_1^M\rangle = \frac{1}{\sqrt{v!}} \left[a_1^\dagger - \frac{\pi^2}{8N}\left(a_1^\dagger \sum_{m'=1}^{M-1} \sin^2 m'\theta_M - a_0^\dagger \sum_{m'=1}^{M-1} \sin m'\theta_M \cos m'\theta_M\right)\right]^v |0,0,0\rangle$$

$$= \frac{1}{\sqrt{v!}} \left\{a_1^\dagger - \frac{\pi^2}{8N}\left[a_1^\dagger\left(-1 + \sum_{m'=1}^{M} \sin^2 m'\theta_M\right) - a_0^\dagger \sum_{m'=1}^{M} \sin m'\theta_M \cos m'\theta_M\right]\right\}^v |0,0,0\rangle. \tag{A19}$$

Regarding these two summations in Eq. (A19), we can do the following approximation,

$$\sum_{m'=1}^{M} \sin^2\left(\frac{m'\pi}{2M}\right) \approx \frac{2M}{\pi} \int_0^{\frac{\pi}{2}} \sin^2 m'\, dm' = \frac{M}{2} \tag{A20}$$

and

$$\sum_{m'=1}^{M} \sin\left(\frac{m'\pi}{2M}\right) \cos\left(\frac{m'\pi}{2M}\right) \approx \frac{2M}{\pi} \int_0^{\frac{\pi}{2}} \sin m' \cos m'\, dm' = \frac{M}{\pi}. \tag{A21}$$

Consequently, Eq. (A19) can be rewritten as

$$|\phi_1^M\rangle = \frac{1}{\sqrt{v!}}\left[\left(1 + \frac{\pi^2}{8N} - \frac{\pi^2 M}{16N}\right) a_1^\dagger + \frac{\pi M}{8N} a_0^\dagger\right]^v |0,0,0\rangle$$

$$\approx \frac{1}{\sqrt{v!}} \left[\frac{\pi M}{8N} a_0^\dagger + \left(1 - \frac{\pi^2 M}{16N}\right) a_1^\dagger\right]^v |0,0,0\rangle. \tag{A22}$$

*This is Eq. (3) in the main text.* In the second line, we use the assumption that $M \gg 1$.

Now we can calculate the probability that $D_1$ and only $D_1$ clicks. Under such condition, it is easy to see that the photon state has to be $|0, v, 0\rangle$, which results in having,

$$P_1 \approx 1 - \frac{\pi^2 v M}{8N}, \tag{A23}$$

which requires that $N \gg Mv$ and $M \gg 1$.

## APPENDIX B: Calculations for coherent state as an input state

We consider an initial coherent photon state

$$|\alpha, 0,0\rangle = \exp(\alpha a_0^\dagger - \alpha^* a_0)|0,0,0\rangle. \tag{B1}$$

Here, we assume that $|\alpha, 0,0\rangle \equiv |\alpha\rangle|0\rangle|0\rangle$ with $|\alpha\rangle = \exp\left(-\frac{1}{2}|\alpha|^2\right)\sum_{v=0}^{\infty}\frac{\alpha^v}{\sqrt{v!}}|v\rangle$ representing the coherent state.

After passing through the first $BS_M$, the photon state is

$$|\Phi_0^1\rangle = |\Phi_1^1\rangle$$
$$= \exp[\alpha(a_0^\dagger \cos\theta_M + a_1^\dagger \sin\theta_M) - \alpha^*(a_0 \cos\theta_M + a_1 \sin\theta_M)]|0,0,0\rangle$$
$$= |\alpha \cos\theta_M, \alpha \sin\theta_M, 0\rangle, \tag{B2}$$

where the superscript of $|\Phi_s^m\rangle$ represents the $m$-th outer interferometer and the subscript represents Bob's signal $s$.

After the first $BS_N$ in the first inner chain, we have

$$\exp\{\alpha[a_0^\dagger \cos\theta_M + (a_1^\dagger \cos\theta_N + a_2^\dagger \sin\theta_N)\sin\theta_M]$$
$$-\alpha^*[a_0 \cos\theta_M + (a_1 \cos\theta_N + a_2 \sin\theta_N)\sin\theta_M]\}|0,0,0\rangle$$
$$= |\alpha \cos\theta_M, \alpha \sin\theta_M \cos\theta_N, \alpha \sin\theta_M \sin\theta_N\rangle. \tag{B3}$$

**B.1 Bob deactivates his detector $D_{2B}$, i.e., his signal is $s = 0$.**

It is not difficult to obtain that after $n$-th $BS_N$, the photon state is

$$\exp\{\alpha[a_0^\dagger \cos\theta_M + (a_1^\dagger \cos n\theta_N + a_2^\dagger \sin n\theta_N)\sin\theta_M]$$
$$-\alpha^*[a_0 \cos\theta_M + (a_1 \cos n\theta_N + a_2 \sin n\theta_N)\sin\theta_M]\}|0,0,0\rangle$$
$$= |\alpha \cos\theta_M, \alpha \sin\theta_M \cos n\theta_N, \alpha \sin\theta_M \sin n\theta_N\rangle. \tag{B4}$$

When $n = N$, we have

$$|\alpha \cos\theta_M, 0, \alpha \sin\theta_M\rangle$$
$$\equiv |\alpha \cos\theta_M\rangle|0\rangle\left[\exp\left(-\frac{1}{2}|\alpha \sin\theta_M|^2\right)\sum_{v=0}^{\infty}\frac{(\alpha \sin\theta_M)^v}{\sqrt{v!}}|v\rangle\right]. \tag{B5}$$

Now, we consider the measurement performed by $D_{2A}$. If the detector does not detect any photons, the photon state in Zone 2 collapse to the zero-photon state, i.e.,

$$\exp\left(-\frac{1}{2}|\alpha \sin\theta_M|^2\right)|\alpha \cos\theta_M\rangle|0\rangle|0\rangle$$
$$= \exp\left[-\frac{1}{2}|\alpha|^2(1 - \cos^2\theta_M)\right]|\alpha \cos\theta_M, 0,0\rangle. \tag{B6}$$

The probability that no photons are detected by $D_{2A}$ is $\exp[-|\alpha|^2(1 - \cos^2\theta_M)]$.

Then, after the second $BS_M$, we have

$$|\Phi_0^2\rangle = \exp\left[-\frac{1}{2}|\alpha|^2(1 - \cos^2\theta_M)\right]|\alpha \cos^2\theta_M, \alpha \cos\theta_M \sin\theta_M, 0\rangle. \tag{B7}$$

After passing through the second inner chain, and if $D_{2A}$ still does not detect any photons, the photon state becomes

$$\exp\left[-\frac{1}{2}|\alpha|^2(1-\cos^2\theta_M) - \frac{1}{2}|\alpha|^2\cos^2\theta_M\sin^2\theta_M\right]|\alpha\cos^2\theta_M, 0, 0\rangle$$

$$= \exp\left[-\frac{1}{2}|\alpha|^2(1-\cos^2\theta_M)(1+\cos^2\theta_M)\right]|\alpha\cos^2\theta_M, 0, 0\rangle$$

$$= \exp\left[-\frac{1}{2}|\alpha|^2(1-\cos^4\theta_M)\right]|\alpha\cos^2\theta_M, 0, 0\rangle. \tag{B8}$$

Then, after the third $BS_M$, we have

$$|\Phi_0^3\rangle = \exp\left[-\frac{1}{2}|\alpha|^2(1-\cos^4\theta_M)\right]|\alpha\cos^3\theta_M, \alpha\cos^2\theta_M\sin\theta_M, 0\rangle. \tag{B9}$$

The above process is repeated many times. After $m$-th $BS_M$, we have

$$|\Phi_0^m\rangle = \exp\left\{-\frac{1}{2}|\alpha|^2\left[1-\cos^{2(m-1)}\theta_M\right]\right\}|\alpha\cos^m\theta_M, \alpha\cos^{m-1}\theta_M\sin\theta_M, 0\rangle. \tag{B10}$$

In addition, after the following inner cycle, if $D_{2A}$ does not click, the photon state becomes

$$\exp\left\{-\frac{1}{2}|\alpha|^2\left[1-\cos^{2(m-1)}\theta_M\right]\right\}\exp\left\{-\frac{1}{2}|\alpha|^2\cos^{2(m-1)}\theta_M\sin^2\theta_M\right\}|\alpha\cos^m\theta_M, 0, 0\rangle$$

$$= \exp\left[-\frac{1}{2}|\alpha|^2(1-\cos^{2m}\theta_M)\right]|\alpha\cos^m\theta_M, 0, 0\rangle. \tag{B11}$$

*This equation can be used to describe the final photon state in the modified scheme for $s = 0$ when $m = m_c$.*

Back to Eq. (B10), when $m = M$, we have

$$|\Phi_0^M\rangle = \exp\left\{-\frac{1}{2}|\alpha|^2\left[1-\cos^{2(M-1)}\theta_M\right]\right\}|\alpha\cos^M\theta_M, \alpha\cos^{M-1}\theta_M\sin\theta_M, 0\rangle$$

$$\approx \exp\left(-\frac{|\alpha|^2\pi^2}{16M}\right)\left|\alpha\left(1-\frac{\pi^2}{8M}\right), \alpha\frac{\pi}{2M}, 0\right\rangle. \tag{B12}$$

In the second line, we have used the approximation

$$\cos^M\theta_M \approx \cos^{M-1}\theta_M \approx 1 - \frac{\pi^2}{8M} \tag{B13}$$

and

$$\sin\theta_M \approx \frac{\pi}{2M}. \tag{B14}$$

We neglect all second or higher order terms of $1/M$.

Now, we can calculate the probability that $D_0$ and only $D_0$ clicks, which is

$$P_0 = \exp\left\{-|\alpha|^2\left[1-\cos^{2(M-1)}\theta_M\right]\right\}\exp\left[-|\alpha|^2\cos^{2(M-1)}\theta_M\sin^2\theta_M\right]$$
$$\times \left[1 - \exp(-|\alpha|^2\cos^{2M}\theta_M)\right]$$
$$= \exp[-|\alpha|^2(1-\cos^{2M}\theta_M)] - \exp(-|\alpha|^2)$$
$$\approx 1 - \frac{|\alpha|^2\pi^2}{4M}. \tag{B15}$$

The term in the second line comes from the requirement that $D_0$ has to detect at least one photon. In addition, to get the third equation, we require that $M \gg |\alpha|^2 \gg 1$.

**B.2 Bob activates his detector $D_{2B}$, i.e., his signal is $s = 1$.**

Based on Eq. (B3), we consider the influence of Bob's measurement. If $D_{2B}$ does not click, we have

$$\exp\left[-\tfrac{1}{2}|\alpha|^2 \sin^2\theta_M (1 - \cos^2\theta_N)\right] |\alpha \cos\theta_M, \alpha \sin\theta_M \cos\theta_N, 0\rangle. \tag{B16}$$

At the end of the first inner chain, if $D_{2B}$ does not click, we have

$$\exp\left[-\tfrac{1}{2}|\alpha|^2 \sin^2\theta_M (1 - \cos^{2N}\theta_N)\right] |\alpha \cos\theta_M, \alpha \sin\theta_M \cos^N\theta_N, 0\rangle$$
$$\approx \exp\left(-\tfrac{\pi}{8N}|\alpha|^2 \sin^2\theta_M\right) \left|\alpha \cos\theta_M, \alpha \sin\theta_M + \left(-\tfrac{\pi^2}{8N}\right)\alpha \sin\theta_M, 0\right\rangle. \tag{B17}$$

In the approximation, we do power series expansion and discard all the second and higher-order terms of $1/N$ due to the assumption $N \gg 1$.

After the second $BS_M$, we have

$$|\Phi_1^2\rangle = \exp\left(-\tfrac{\pi^2}{8N}|\alpha|^2 \sin^2\theta_M\right)$$
$$\times \left|\alpha \cos 2\theta_M + \tfrac{\pi^2}{8N}\alpha \sin\theta_M \sin\theta_M, \alpha \sin 2\theta_M - \tfrac{\pi^2}{8N}\sin\theta_M \cos\theta_M, 0\right\rangle. \tag{B18}$$

The above process is repeated in the next outer interferometer. With the same approximation, after the third $BS_M$, we have

$$|\Phi_1^3\rangle = \exp\left(-\tfrac{\pi^2}{8N}|\alpha|^2 \sin^2\theta_M\right) \exp\left(-\tfrac{\pi^2}{8N}|\alpha|^2 \sin^2 2\theta_M\right)$$
$$\times \left|\alpha \cos 3\theta_M + \tfrac{\pi^2}{8N}\alpha \sin\theta_M \sin 2\theta_M + \tfrac{\pi^2}{8N}\alpha \sin 2\theta_M \sin\theta_M\right\rangle$$
$$\otimes \left|\alpha \sin 3\theta_M - \tfrac{\pi^2}{8N}\alpha \sin 2\theta_M \cos\theta_M - \tfrac{\pi^2}{8N}\alpha \sin\theta_M \cos 2\theta_M\right\rangle |0\rangle. \tag{B19}$$

Similarly, after $m$-th $BS_M$, if $D_{2B}$ does not click, we have

$$|\Phi_1^m\rangle = \exp\left(-\tfrac{\pi^2}{8N}|\alpha|^2 \sum_{m'=1}^{m-1} \sin^2 m'\theta_M\right)$$
$$\times \left|\alpha \cos m\theta_M + \tfrac{\pi^2}{8N}\alpha \sum_{m'=1}^{m-1} \sin m'\theta_M \sin(m - m')\theta_M\right\rangle$$
$$\otimes \left|\alpha \sin m\theta_M - \tfrac{\pi^2}{8N}\alpha \sum_{m'=1}^{m-1} \sin m'\theta_M \cos(m - m')\theta_M\right\rangle |0\rangle. \tag{B20}$$

In addition, after the following inner chain, if no $D_{2B}$ clicks, we have

$$\exp\left(-\tfrac{\pi^2}{8N}|\alpha|^2 \sum_{m'=1}^{m-1} \sin^2 m'\theta_M\right) \exp\left(-\tfrac{1}{2}\left|\tfrac{\pi}{2N}\alpha \sin m\theta_M\right|^2\right)$$
$$\times \left|\alpha \cos m\theta_M + \tfrac{\pi^2}{8N}\alpha \sum_{m'=1}^{m-1} \sin m'\theta_M \sin(m - m')\theta_M\right\rangle$$

$$\otimes \left| \alpha \sin m\theta_M - \frac{\pi^2}{8N} \alpha \sum_{m'=1}^{m} \sin m'\theta_M \cos(m-m')\theta_M \right\rangle |0\rangle$$

$$= \exp\left(-\frac{\pi^2}{8N}|\alpha|^2 \sum_{m'=1}^{m} \sin^2 m'\theta_M\right)$$

$$\times \left| \alpha \cos m\theta_M + \frac{\pi^2}{8N} \alpha \sum_{m'=1}^{m-1} \sin m'\theta_M \sin(m-m')\theta_M \right\rangle$$

$$\otimes \left| \alpha \sin m\theta_M - \frac{\pi^2}{8N} \alpha \sum_{m'=1}^{m} \sin m'\theta_M \cos(m-m')\theta_M \right\rangle |0\rangle. \quad (B21)$$

*This equation can be used to describe the final photon state in the modified scheme for $s = 1$ when $m = m_c$.*

Back to Eq. (B20), when $m = M$, we have

$$|\Phi_1^M\rangle = \exp\left(-\frac{\pi^2}{8N}|\alpha|^2 \sum_{m'=1}^{M-1} \sin^2 m'\theta_M\right)$$

$$\times \left| \frac{\pi^2}{8N} \alpha \sum_{m'=1}^{M-1} \sin m'\theta_M \sin(M-m')\theta_M, \alpha\left[1 - \frac{\pi^2}{8N} \sum_{m'=1}^{M-1} \sin m'\theta_M \cos(M-m')\theta_M\right], 0 \right\rangle$$

$$= \exp\left[-\frac{\pi^2}{8N}|\alpha|^2\left(-1 + \sum_{m'=1}^{M} \sin^2 m'\theta_M\right)\right]$$

$$\times \left| \frac{\pi^2}{8N} \alpha \sum_{m'=1}^{M} \sin m'\theta_M \cos m'\theta_M, \alpha\left[1 - \frac{\pi^2}{8N}\left(-1 + \sum_{m'=1}^{M} \sin^2 m'\theta_M\right)\right], 0 \right\rangle. \quad (B22)$$

Utilizing the approximation shown in Eqs. (A20) and (A21), we have

$$|\Phi_1^M\rangle = \exp\left[-|\alpha|^2\left(-\frac{\pi^2}{8N} + \frac{\pi^2 M}{16N}\right)\right] \left| \alpha \frac{\pi M}{8N}, \alpha\left(1 + \frac{\pi^2}{8N} - \frac{\pi^2 M}{16N}\right), 0 \right\rangle$$

$$\approx \exp\left(-|\alpha|^2 \frac{\pi^2 M}{16N}\right) \left| \alpha \frac{\pi M}{8N}, \alpha\left(1 - \frac{\pi^2 M}{16N}\right), 0 \right\rangle. \quad (B23)$$

Now, we can calculate the probability that $D_1$ and only $D_1$ clicks, which is

$$P_1 = \exp\left(-|\alpha|^2 \frac{\pi^2 M}{8N}\right) \exp\left(-|\alpha|^2 \frac{\pi^2 M^2}{64N^2}\right) \left\{1 - \exp\left[-|\alpha|^2 \left(1 - \frac{\pi^2 M}{16N}\right)^2\right]\right\}$$

$$\approx \exp\left(-|\alpha|^2 \frac{\pi^2 M}{8N}\right) - \exp(-|\alpha|^2)$$

$$\approx 1 - |\alpha|^2 \frac{\pi^2 M}{8N}. \quad (B24)$$

In the first line of Eq. (B24), the first factor $\exp(-|\alpha|^2 \pi^2 M/8N)$ is derived from the probability amplitude in Eq. (B23). The second factor $\exp(-|\alpha|^2 \pi^2 M^2/64N^2)$ is due to the condition that $D_0$ does not click. The third factor $\{1 - \exp[-|\alpha|^2(1 - \pi^2 M/16N)^2]\}$ is due to the condition that $D_1$ detects at least one photon. In the second line of Eq. (B24), we have neglected all second order terms of $M/N$ since $N \gg M$. In the third line of Eq. (B24), we use $\exp(-|\alpha|^2) \approx 0$ since we assume $|\alpha|^2 \gg 1$, and $N \gg |\alpha|^2 M$. Combined with the requirement obtained in (B15), *the condition for counterfactual communication is $N \gg |\alpha|^2 M \gg |\alpha|^4$.*

### APPENDIX C: Calculations for the input state as arbitrary photon statistics

Consider an arbitrary photon statistics input, which is

$$\sum_{v=0}^{\infty} c_v |v,0,0\rangle. \tag{C1}$$

**C.1 Bob deactivates his detector $D_{2B}$, i.e., his signal is $s = 0$.**

Utilizing Eq. (A9), we have

$$|\Psi_0^M\rangle = \sum_{v=0}^{\infty} \frac{c_v}{\sqrt{v!}} \left[\cos^{(M-1)} \theta_M \left(a_0^\dagger \cos \theta_M + a_1^\dagger \sin \theta_M\right)\right]^v |0,0,0\rangle$$

$$\approx \sum_{v=0}^{\infty} \frac{c_v}{\sqrt{v!}} \left[a_0^\dagger \left(1 - \frac{\pi^2}{8M}\right) + a_1^\dagger \frac{\pi}{2M}\right]^v |0,0,0\rangle. \tag{C2}$$

Then, the probability that $D_0$ and only $D_0$ clicks is

$$P_0 = \sum_{v=1}^{\infty} |c_v|^2 \cos^{2Mv} \theta_M \approx \sum_{v=1}^{\infty} |c_v|^2 \left(1 - \frac{\pi^2}{8M}\right)^{2v}. \tag{C3}$$

When considering light filed in real life scenario, the average photon number is limited. Therefore, we can assume a cut-off photon number $v_c$ such that the Fock states with $v > v_c$ have no contribution to the average photon number $\bar{v} = \sum_{v=0}^{\infty} |c_v|^2 v$, i.e.,

$$\sum_{v=v_c}^{\infty} |c_v|^2 v = 0. \tag{C4}$$

Due to $\sum_{v=v_c}^{\infty} |c_v|^2 v \geq \sum_{v=v_c}^{\infty} |c_v|^2$, we have $\sum_{v=v_c}^{\infty} |c_v|^2 = 0$. Consequently, when $M \gg v_c$, we can obtain

$$P_0 \approx \sum_{v=1}^{v_c} |c_v|^2 \left(1 - \frac{\pi^2}{8M}\right)^{2v} = \sum_{v=1}^{v_c} |c_v|^2 \left(1 - \frac{v\pi^2}{4M}\right) = 1 - \frac{\pi^2}{4M} \bar{v} - |c_0|^2. \tag{C5}$$

We emphasize that the above condition may be changed when we consider a specific photon statistic such as a coherent state.

Considering the coherent state $|\alpha\rangle = \exp\left(-\frac{1}{2}|\alpha|^2\right) \sum_{v=0}^{\infty} \frac{\alpha^v}{\sqrt{v!}} |v\rangle$, based on Eq. (C3), the corresponding probability is

$$P_0 = -|c_0|^2 + \sum_{v=0}^{\infty} |c_v|^2 \cos^{2Mv} \theta_M$$

$$= -\exp(-|\alpha|^2) + \sum_{v=0}^{\infty} \exp(-|\alpha|^2) \frac{|\alpha|^{2v}}{v!} \cos^{2Mv} \theta_M$$

$$= -\exp(-|\alpha|^2) + \sum_{v=0}^{\infty} \exp[-|\alpha|^2(1 - \cos^{2M} \theta_M + \cos^{2M} \theta_M)] \frac{|\alpha \cos^M \theta_M|^{2v}}{v!}$$

$$= -\exp(-|\alpha|^2) + \exp[-|\alpha|^2 (1 - \cos^{2M} \theta_M)] \sum_{v=0}^{\infty} \exp(-|\alpha \cos^M \theta_M|^2) \frac{|\alpha \cos^M \theta_M|^{2v}}{v!}$$
$$= -\exp(-|\alpha|^2) + \exp[-|\alpha|^2 (1 - \cos^{2M} \theta_M)], \tag{C6}$$

which is in agreement with Eq. (B15).

**C.2 Bob activates his detector $D_{2B}$, i.e., his signal is $s = 1$.**

Utilizing Eq. (A22), we have

$$|\Psi_1^M\rangle = \sum_{v=0}^{\infty} \frac{c_v}{\sqrt{v!}} \left[ \frac{\pi M}{8N} a_0^\dagger + \left(1 - \frac{\pi^2 M}{16N}\right) a_1^\dagger \right]^v |0,0,0\rangle. \tag{C7}$$

Then, the probability that $D_1$ and only $D_1$ clicks is

$$P_1 = \sum_{v=1}^{\infty} |c_v|^2 \left(1 - \frac{\pi^2 M}{16N}\right)^{2v}. \tag{C8}$$

Based on the approximation shown in Eq. (C4), when $N \gg Mv_c$, we have

$$P_1 \approx \sum_{v=1}^{v_c} |c_v|^2 \left(1 - \frac{\pi^2 M}{16N}\right)^{2v} = 1 - \frac{\pi^2 M}{8N} \bar{v} - |c_0|^2. \tag{C9}$$

*When $|c_0|^2 \to 0$, Eqs. (C5) and (C9) are Eq. (5) in the main text.*

Next, we consider the coherent state. Based on Eq. (C8), the corresponding probability is

$$\begin{aligned} P_1 &= -|c_0|^2 + \sum_{v=0}^{\infty} |c_v|^2 \left(1 - \frac{\pi^2 M}{16N}\right)^{2v} \\ &= -\exp(-|\alpha|^2) + \exp\left[-|\alpha|^2 \left(\frac{\pi^2 M}{8N} - \frac{\pi^4 M^2}{16^2 N^2}\right)\right] \\ &\quad \times \sum_{v=0}^{\infty} \exp\left[-|\alpha|^2 \left(1 - \frac{\pi^2 M}{16N}\right)^2\right] \frac{1}{v!} \left|\alpha \left(1 - \frac{\pi^2 M}{16N}\right)\right|^{2v} \\ &= -\exp(-|\alpha|^2) + \exp\left[-|\alpha|^2 \left(\frac{\pi^2 M}{8N} - \frac{\pi^4 M^2}{16^2 N^2}\right)\right] \\ &\approx 1 - \frac{|\alpha|^2 \pi^2 M}{8N} - \exp(-|\alpha|^2), \end{aligned} \tag{C10}$$

which is in agreement with Eq. (B24).

### APPENDIX D: Calculations for modified scheme for strong coherent state input

In this calculation, we mainly compare between the SLAZ scheme and the modified scheme using the same light source. As a frame of reference, we assume that the cycle numbers of the SLAZ protocol using a single photon source are $N'$ and $M'$, respectively. Accordingly, the probability of only $D_0$ or $D_1$ clicking are

$$P_0' = 1 - \frac{\pi^2}{4M'}, \quad P_1' = 1 - \frac{\pi^2 M'}{8N'}. \tag{D1}$$

**D.1 Original SLAZ scheme with strong coherent input**

Apparently, due to Eq. (B15) and Eq. (B24), when $|\alpha|^2$ is large, the condition for $P_0 = P_0'$ and $P_1 = P_1'$ are

$$M = |\alpha|^2 M', \tag{D2}$$

$$N = \frac{|\alpha|^2 N' M}{M'} = |\alpha|^4 N'. \tag{D3}$$

Therefore, the total cycle number is

$$MN = |\alpha|^6 M'N'. \tag{D4}$$

**D.2 Modified scheme for strong coherent input**

Following the discussion in the main text, we define that $p_s$ is the probability that $D_{2A}$ and $D_{2B}$ do not click, which describes the counterfactuality of the communication process, where $s = 0,1$. In addition, we define that $f_s$ is the probability of at least one photon being detected in Zone $s$. Then, $\tilde{P}_s = f_s p_s$.

According to Eqs. (B11) and (B21), which can describe the state of photons after $m_c$-th $BS_M$ and $m_c$-th inner chain for $s = 0,1$, respectively, we have

$$f_0 = 1 - \exp(-|\alpha|^2 \cos^{2m_c} \theta_M), \tag{D5}$$

$$p_0 = \exp[-|\alpha|^2 (1 - \cos^{2m_c} \theta_M)], \tag{D6}$$

$$f_1 = 1 - \exp\left[-\left|\alpha \sin m_c \theta_M - \frac{\pi^2}{8N} \alpha \sum_{m'=1}^{m_c} \sin m'\theta_M \cos(m_c - m')\theta_M\right|^2\right], \tag{D7}$$

$$p_1 = \exp\left(-\frac{\pi^2}{4N} |\alpha|^2 \sum_{m'=1}^{m_c} \sin^2 m'\theta_M\right). \tag{D8}$$

In Eq. (D7), $\left|\alpha \sin m_c \theta_M - \frac{\pi^2}{8N} \alpha \sum_{m'=1}^{m_c} \sin m'\theta_M \cos(m_c - m')\theta_M\right|^2$ in the exponential term represents the average photon number in Zone 1. In the ideal case, this number is expected to be

$$\bar{k} = |\alpha|^2 \sin^2 m_c \theta_M. \tag{D9}$$

When $\bar{k} \ll |\alpha|^2$, the approximation $\sin m_c \theta_M \approx m_c \theta_M$ is valid. Based on this approximation, we calculate Eqs. (D5)-(D8) in the following.

We first calculate Eq. (D8),

$$p_1 = \exp\left(-\frac{\pi^2}{4N} |\alpha|^2 \sum_{m'=1}^{m_c} \sin^2 m'\theta_M\right)$$

$$\approx \exp\left(-\frac{\pi^4}{16NM^2} |\alpha|^2 \sum_{m'=1}^{m_c} m'^2\right)$$

$$= \exp\left[-\frac{\pi^4}{96NM^2} |\alpha|^2 m_c (m_c + 1)(2m_c + 1)\right]. \tag{D10}$$

In the calculation, we have used

$$\sum_{m'=1}^{m_c} m'^2 = \frac{1}{6} m_c (m_c + 1)(2m_c + 1). \tag{D11}$$

We next calculate Eq. (D7). Here we focus on its power exponent part

$$\left|\alpha \sin m_c \theta_M - \frac{\pi^2}{8N} \alpha \sum_{m'=1}^{m_c} \sin m'\theta_M \cos(m_c - m')\theta_M\right|^2$$

$$= \left|\alpha \sin m_c \theta_M - \frac{\pi^2}{8N} \alpha \sum_{m'=1}^{m_c} \sin m'\theta_M (\cos m_c \theta_M \cos m'\theta_M + \sin m_c \theta_M \sin m'\theta_M)\right|^2.$$

$$\tag{D12}$$

Since $m_c\theta_M$ is small and $m' < m_c$, we can neglect the second and higher order terms of $m_c\theta_M$ and $m'\theta_M$. Then, Eq. (D12) can be approximated as

$$\approx \left|\alpha m_c\theta_M - \frac{\pi^2}{8N}\alpha \sum_{m'=1}^{m_c} m'\theta_M\left[\left(1 - \frac{(m_c\theta_M)^2}{2}\right)\left(1 - \frac{(m'\theta_M)^2}{2}\right) + (m_c\theta_M)(m'\theta_M)\right]\right|^2 \quad (D13)$$

$$\approx \left|\alpha m_c\theta_M - \frac{\pi^2}{8N}\alpha\theta_M \sum_{m'=1}^{m_c} m'\right|^2$$

$$= \left(\frac{|\alpha|m_c\pi}{2M}\right)^2 \left[1 - \frac{\pi^2}{16N}(1 + m_c)\right]^2. \quad (D14)$$

As a result, $f_1$ can be rewritten as

$$f_1 = 1 - \exp\left\{-\left(\frac{|\alpha|m_c\pi}{2M}\right)^2\left[1 - \frac{\pi^2}{16N}(1+m_c)\right]^2\right\}. \quad (D15)$$

Accordingly, the expression of $\tilde{P}_1$ is

$$\tilde{P}_1 = \exp\left[-\frac{|\alpha|^2\pi^4 m_c(m_c+1)(2m_c+1)}{96M^2N}\right]\left\{1 - \exp\left[-\frac{|\alpha|^2\pi^2 m_c^2}{4M^2}\left(1 - \pi^2\frac{1+m_c}{16N}\right)^2\right]\right\}. \quad (D16)$$

Similarly, we can calculate $\tilde{P}_0$, which is

$$\tilde{P}_0 = [1 - \exp(-|\alpha|^2\cos^{2m_c}\theta_M)]\exp[-|\alpha|^2(1-\cos^{2m_c}\theta_M)]$$
$$= \exp[-|\alpha|^2(1-\cos^{2m_c}\theta_M)] - \exp(-|\alpha|^2)$$
$$\approx \exp(-|\alpha|^2 m_c\theta_M^2) = \exp\left(-|\alpha|^2\frac{m_c\pi^2}{4M^2}\right). \quad (D17)$$

*This is Eq. (6) in the main text.* In the calculation of Eq. (D17), we have assumed that $\exp(-|\alpha|^2) \approx 0$ due to the strong input.

Above we have given the expressions for $\tilde{P}_0$ and $\tilde{P}_1$. However, it is often necessary to estimate the parameters $M$, $N$ and $m_c$ according to the given $\tilde{P}_0$, $\tilde{P}_1$ and $|\alpha|^2$. To achieve this goal, we need further approximation.

First, we convert the form of Eq. (D17) as follows.

$$M = \sqrt{-\frac{|\alpha|^2 m_c\pi^2}{4\ln\tilde{P}_0}} \quad (D18)$$

We mention that $M$ determines the transmittance of $BS_M$, and we can see that $M$ depends on $m_c$.

By substituting Eq. (D18) into Eq. (D16), we have

$$\tilde{P}_1 = \exp\left[\frac{\pi^2(m_c+1)(2m_c+1)\ln\tilde{P}_0}{24N}\right]\left\{1 - \exp\left[m_c\ln\tilde{P}_0\left(1 - \pi^2\frac{1+m_c}{16N}\right)^2\right]\right\}. \quad (D19)$$

*This is Eq. (7) in the main text.* We notice that the first exponential term $\exp[\pi^2(m_c+1)(2m_c+1)\ln\tilde{P}_0/24N]$ is coming from $p_1$. In order to ensure that the probability of detecting photons in the transmission channel is small, i.e., $p_1 \to 1$, we require that $N \gg m_c^2$. Based on this assumption, we can rewrite Eq. (D19) by doing power series expansion according to $1/N$ and omitting the second and higher order terms of $1/N$. For example, the second exponential term in Eq. (D19) can be approximately rewritten as

$$\exp\left[m_c \ln \tilde{P}_0 \left(1 - \pi^2 \frac{1+m_c}{16N}\right)^2\right]$$

$$\approx \exp\left[m_c \ln \tilde{P}_0 \left(1 - \pi^2 \frac{1+m_c}{8N}\right)\right]$$

$$= \exp\left[-\frac{\pi^2}{8N}(1+m_c)m_c \ln \tilde{P}_0\right]\exp(m_c \ln \tilde{P}_0)$$

$$\approx \left[1 - \frac{\pi^2}{8N}m_c(1+m_c)\ln \tilde{P}_0\right]\exp(m_c \ln \tilde{P}_0). \tag{D20}$$

Here, we mention that $\bar{k} = |\alpha|^2 \sin^2 m_c \theta_M \approx |\alpha|^2 m_c^2 \theta_M^2$. In addition, according to Eq. (D17), we have $\ln \tilde{P}_0 = -|\alpha|^2 m_c \theta_M^2$. Therefore, it is easy to see that

$$\bar{k} = -m_c \ln \tilde{P}_0 . \tag{D21}$$

Since $\bar{k}$ represents the average number of photons appearing in Zone 1, the larger its value, the higher the probability that Alice can successfully distinguish Bob's signals. Therefore, the value of $\bar{k}$ can be greater than 1, and we need to keep the term $\exp(m_c \ln \tilde{P}_0)$ in the calculation.

Regarding $\tilde{P}_1$, similarly, we can obtain

$$\tilde{P}_1 \approx \left[1 + \frac{\pi^2(m_c+1)(2m_c+1)\ln \tilde{P}_0}{24N}\right]\left\{1 - \left[1 - \frac{\pi^2}{8N}m_c(1+m_c)\ln \tilde{P}_0\right]\exp(m_c \ln \tilde{P}_0)\right\}$$

$$\approx 1 - \left[1 - \frac{\pi^2}{8N}m_c(1+m_c)\ln \tilde{P}_0\right]\exp(m_c \ln \tilde{P}_0) + \frac{\pi^2}{24N}(m_c+1)(2m_c+1)\ln \tilde{P}_0$$

$$- \frac{\pi^2}{24N}(m_c+1)(2m_c+1)\ln \tilde{P}_0 \exp(m_c \ln \tilde{P}_0)$$

$$= 1 + \frac{\pi^2}{24N}(m_c+1)(2m_c+1)\ln \tilde{P}_0 - \exp(m_c \ln \tilde{P}_0)\left[1 - \frac{\pi^2}{24N}(m_c^2-1)\ln \tilde{P}_0\right]. \tag{D22}$$

Based on Eq. (D22), we can give the expression of $N$, which is

$$N = \frac{\pi^2(m_c+1)[(2m_c+1)+(m_c-1)\exp(m_c \ln \tilde{P}_0)]\ln \tilde{P}_0}{24[\tilde{P}_1 + \exp(m_c \ln \tilde{P}_0) - 1]}. \tag{D23}$$

*This is Eq. (8) in the main text.*

Regarding the total cycle number $T = Nm_c$, we have

$$T = \frac{\pi^2 m_c(m_c+1)[(2m_c+1)+(m_c-1)\exp(m_c \ln \tilde{P}_0)]\ln \tilde{P}_0}{24[\tilde{P}_1 + \exp(m_c \ln \tilde{P}_0) - 1]}. \tag{D24}$$

Eq. (D24) indicates that we can scan $m_c$ to get the minimum $T$. When $m_c$ is determined, based on Eqs. (D18) and (D23), we can get the corresponding $M$ and $N$. *In the main text, the red solid curve in Fig.1(d) is drawn in this way.* Regarding the circle points in the figure, we scan $M$, $N$ and $m_c$ without any approximation.

Based on Eq. (D24), we prove that the modified scheme requires much less resources than the original scheme using the same light source. We utilize Eq. (D21) to rewrite Eq. (D24), which leads to

$$T = \frac{\pi^2 \bar{k}(\ln \tilde{P}_0 - \bar{k})[2\bar{k} - \ln \tilde{P}_0 + (\bar{k} + \ln \tilde{P}_0) \exp(-\bar{k})]}{24(\ln \tilde{P}_0)^2 [\exp(-\bar{k}) - (1 - \tilde{P}_1)]}. \tag{D25}$$

Eq. (D25) describes the relation between the resources ($T$) required by the modified scheme and the average number of photons expected to be detected in Zone 1 ($\bar{k}$). In particular, when $\bar{k}$ is very large, the probability that $D_1$ does not find any photons is negligible, i.e., $\exp(-\bar{k}) \approx 0$ (When the average photon number $|\alpha|^2$ of the light source is large, this condition is not difficult to achieve, while ensuring $\bar{k} \ll |\alpha|^2$. For example, when $\bar{k} = 5$, $\exp(-\bar{k}) \approx 0.007$), we can obtain

$$T \approx \frac{\pi^2 \bar{k}^3}{12(\ln \tilde{P}_0)^2 (1 - \tilde{P}_1)}. \tag{D26}$$

By substituting (D1) into (D26), i.e., $\tilde{P}_s = P_s'$, we have

$$T \approx \frac{32}{3\pi^4} \bar{k}^3 M' N' \approx \frac{1}{10} \bar{k}^3 M' N'. \tag{D27}$$

Obviously, due to $\bar{k} \ll |\alpha|^2$, the value given in Eq. (D27) is much smaller than that is given by Eq. (D4). Moreover, here $T$ is not necessarily the minimum. Therefore, the modified scheme can save resources.

*So far, we have obtained all the results shown in the main text.* Next, we discuss another situation, in which, we only consider the probability that the photon state at Bob's end is a vacuum state ($p_s$) and ignore Alice's detectors receiving the photon or not. Based on the assumptions that $m_c \theta_M$ is small and $p_s$ is close to 1, we have

$$p_0 = \exp\left(-|\alpha|^2 \frac{m_c \pi^2}{4M^2}\right) = \exp\left(-\frac{\bar{k}}{m_c}\right), \tag{D28}$$

where we have used the relation $\bar{k} \approx |\alpha|^2 m_c^2 \pi^2 / 4M^2$, which leads to

$$m_c = -\frac{\bar{k}}{\ln p_0}. \tag{D29}$$

Eq. (D28) also indicates that

$$M = \sqrt{-\frac{|\alpha|^2 m_c \pi^2}{4 \ln p_0}}. \tag{D30}$$

Regarding $p_1$, according to Eq. (D10), we can approximately obtain,

$$\begin{aligned} p_1 &= \exp\left[-\frac{\pi^4}{96NM^2}|\alpha|^2 m_c(m_c + 1)(2m_c + 1)\right] \\ &\approx \exp\left(-\frac{\pi^4}{48NM^2}|\alpha|^2 m_c^3\right) \\ &= \exp\left(\frac{\pi^2 \bar{k}^2}{12N \ln p_0}\right). \end{aligned} \tag{D31}$$

In the second line, we have assumed that $m_c \gg 1$ for simplicity. Eq. (D31) also leads to,

$$N = \frac{\pi^2 \bar{k}^2}{12 \ln p_0 \ln p_1}. \tag{D32}$$

Then, the resources required in the current case is,

$$T = -\frac{\pi^2 \bar{k}^3}{12 (\ln p_0)^2 \ln p_1}. \tag{D33}$$

Compared with the single-photon case, assuming $p_s = P'_s = P'$, we have

$$\log_{10} T = \log_{10} \left[ -\frac{\pi^2 \bar{k}^3}{12 (\ln P')^3} \right]. \tag{D34}$$

In addition, for a given $\bar{k}$, we can rewrite Eqs.(D29)(D30)(D32)(D33) by using Eq. (D1), which leads to,

$$m_c \approx \frac{4 M' \bar{k}}{\pi^2}, \tag{D35}$$

$$M \approx \frac{2 M' \sqrt{\bar{k}} |\alpha|}{\pi}, \tag{D36}$$

$$N \approx \frac{8 N' \bar{k}^2}{3 \pi^2}, \tag{D37}$$

$$T = N m_c = \frac{32}{3 \pi^4} \bar{k}^3 M' N' \approx \frac{1}{10} \bar{k}^3 M' N'. \tag{D38}$$

It is worth noting that Eqs. (D38) and (D27) are the same.

In order to verify our conclusion, we plot Fig. D1. The black dashed curve is plotted for Eq. (D34) with $\bar{k} = 2$. The blue dotted curve is plotted for the single photon case, in which we scan $M'$ and $N'$ for minimum $\log_{10} T$ corresponding to $P'$, where $T = M'N'$. We obtain this result without using any approximation. As we expected, the black dashed curve and the blue dotted curve almost coincide. When $P'$ is small, the blue dotted curve starts off in steps-shape fashion. The reason is that $M'$ and $N'$ are integers in the simulation. According to Eq. (D1), as $M'$ increases, $N'$ must also increase subsequently to ensure that $P'$ does not decrease. In the meantime, when $P'$ is small, the required $M'$ and $N'$ are also small, which leads to more obvious impact on $T$. These facts lead to the steps-shape seen in the figure. As $P'$ increases, the corresponding $M'$ and $N'$ also increase resulting in smooth blue dotted curve without further steps. In addition, for comparison, the red solid curve is plotted for minimum $\log_{10} T$ according to Eq. 8 in the main text by setting $\tilde{P}_s = P'$ (i.e., the red solid curve in Fig.1(d)), which takes into account the probability that Alice needs to receive at least one photon. Fig. D1 indicates that when using a multiphoton light source, a significant increase in resources is not a necessary condition to ensure that no photons enter the transmission channel. The real benefit of increasing system resources is to improve the efficiency of information transmission. In short, this is because $\bar{k}$ is proportional to the trigger probability of the Alice's detector. As $\bar{k}$ increases, $T$ also increases according to Eq. (D38).

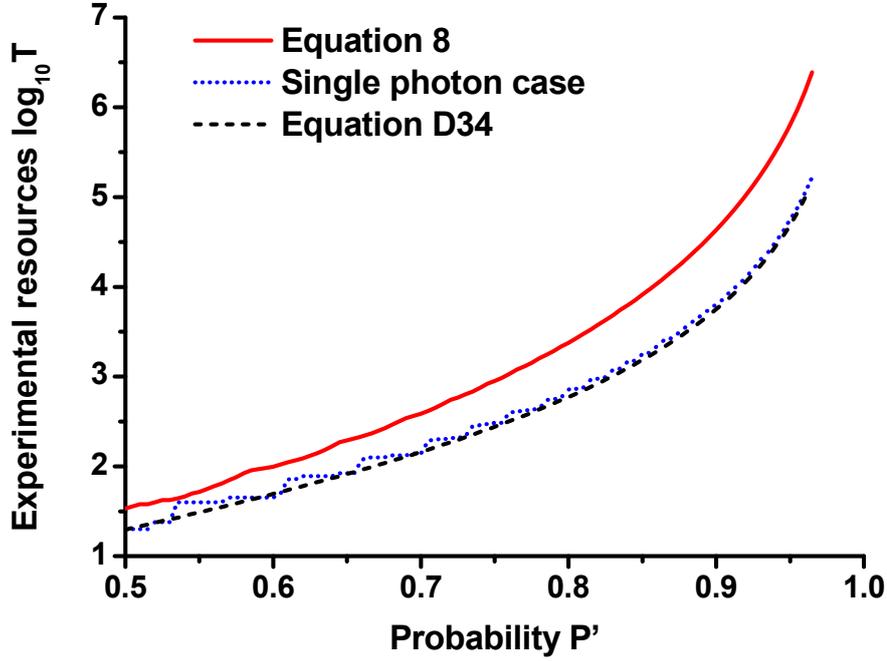

Fig. D1. Comparison of resources. The blue dotted curve is plotted for the original SLAZ protocol using the single photon source. The other two curves are plotted for the modified scheme. The red solid curve is based on Eq. 8 in the main text with $|\alpha|^2 = 200$ (i.e., the red solid curve in Fig.1(d)). The black dashed curve is plotted for Eq. (D34) with $\bar{k} = 2$. The difference is that for the black dashed curve, $P'$ is the probability that no photon appears in the public transmission channel whether or not Alice's detectors receive the photon. The blue dotted and black dashed curves are almost overlapped. The steps-shape of the blue dotted curve is the result of the $M'$ and $N'$ with integer values in the simulation.

**Reference:**
[1] H. Salih, Z.-H. Li, M. Al-Amri, and M. S. Zubairy, Phys. Rev. Lett. **110**, 170502 (2013).
[2] R. H. Dicke, Am. J. Phys. **49**, 925–930 (1981).
[3] A. C. Elitzur, and L. Vaidman, Found. Phys. **23**, 987 (1993).
[4] P. G. Kwiat, H. Weinfurter, T. Herzog, A. Zeilinger, and M. A. Kasevich, Phys. Rev. Lett. **74**, 4763 (1995).
[5] O. Hosten, M. T. Rakher, J. T. Barreiro, N. A. Peters, and P. G. Kwiat, Nature **439**, 949 (2006).
[6] Y. Aharonov, and D. Rohrlich, Phys. Rev. Lett. **125**, 260401 (2020).
[7] J. R. Hance, J. Ladyman, and J. Rarity, Found. Phys. **51**, 12 (2021).
[8] R. B. Griffiths, Phys. Rev. A **94**, 032115 (2016).
[9] Z.-H. Li, M. Al-Amri, and M. S. Zubairy, Phys. Rev. A **88**, 046102 (2013).
[10] L. Vaidman, Phys. Rev. Lett. **98**, 160403 (2007).
[11] L. Vaidman, Phys. Rev. A **87**, 052104 (2013).

[12] N. Gisin, Phys. Rev. A **88**, 030301(R) (2013).
[13] D. R. M. Arvidsson-Shukur and C. H. W. Barnes, Phys. Rev. A **94**, 062303 (2016).
[14] D. R. M. Arvidsson-Shukur, A. N. O. Gottfries, and C. H. W. Barnes, Phys. Rev. A **96**, 062316 (2017).
[15] I. Alonso Calafell, T. Strömberg, D. R. M. Arvidsson-Shukur, L. A. Rozema, V. Saggio, C. Greganti, N. C. Harris, M. Prabhu, J. Carolan, M. Hochberg, T. Baehr-Jones, D. Englund, C. H. W. Barnes and P. Walther, npj Quantum Information **61** (2019).
[16] Y. Aharonov and L. Vaidman, Phys. Rev. A **99**, 010103(R) (2019).
[17] D. R. M. Arvidsson-Shukur and C. H. W. Barnes, Phys. Rev. A **99**, 060102(R) (2019).
[18] Q. Guo, L.-Y. Cheng, L. Chen, H.-F. Wang, and S. Zhang, Opt. Express **22**, 8970 (2014).
[19] Y. Y. Chen, X. M. Gu, D. Jiang, L. Xie, and L. J. Chen, Opt. Express **23**, 21193 (2015).
[20] Y. Y. Chen, D. Jiang, X. M. Gu, L. Xie, and L. J. Chen, J. Opt. Soc. Am. B **33**, 663 (2016).
[21] Q. Guo, L.-Y. Cheng, H.-F. Wang, and S. Zhang, Opt. Express **26**, 27314 (2018).
[22] Q. Guo, L.-Y. Cheng, L. Chen, H.-F. Wang, and S. Zhang, Sci. Rep. **5**, 8416 (2015).
[23] Z.-H. Li, M. Al-Amri, and M. S. Zubairy, Phys. Rev. A **92**, 052315 (2015).
[24] H. Salih, Front. Phys. **3**, 94 (2016).
[25] Z.-H. Li, M. Al-Amri, X. H. Yang, and M. S. Zubairy, Phys. Rev. A **100**, 022110 (2019).
[26] Z.-H. Li, X.-F. Ji, S. Asiri, L.-J. Wang, and M. Al-Amri, Phys. Rev. A **102**, 022606 (2020).
[27] Q. Guo, L.-Y. Cheng, L. Chen, H.-F. Wang, and S. Zhang, Phys. Rev. A **90**, 042327 (2014).
[28] Z. Cao, Phys. Rev. A **102**, 052413 (2020).
[29] F. Zaman, Y. Jeong, and H. Shin, Sci. Rep. **8**, 14641 (2018).
[30] J. R. Hance and J. Rarity, npj Quantum Information **7**, 88 (2021).
[31] Q. Guo, S. Zhai, L.-Y. Cheng, H.-F. Wang, and S. Zhang, Phys. Rev. A **96**, 052335 (2017).
[32] C. Liu, X. F. Yang, L. J. Cui, S. M. Zhou, and J. X. Zhang, Opt. Express **28**, 21916 (2020).
[33] Y. Cao, Y.-H. Li, Z. Cao, J. Yin, Y.-A. Chen, H.-L. Yin, T.-Y. Chen, X. F. Ma, C.-Z. Peng, and J.-W. Pan, Proc. Natl. Acad. Sci. USA **114**, 4920 (2017).
[34] C. Liu, J.-H. Liu, J.-X. Zhang, and S.-Y. Zhu, Sci. Rep. **7**, 10875 (2017).
[35] L. J. Wang, Z.-H. Li, J. P. Xu, Y. P. Yang, M. Al-Amri, and M. S. Zubairy, Opt. Express **27**, 20525 (2019).